\documentclass[twocolumn]{revtex4-1}
\begin{document}
{\bf Criticism of ``Asking Photons Where They Have Been''}

In a recent work \cite{DANAN}, Danan, Farfurnik, Bar-Ad, and Vaidman claim to have provided ``suprizing experimental evidence regarding the past of the photons''. The Letter was widely celebrated, being chosen both as an Editor's suggesstion and for the Viewpoint in {\em Physics}, and is accepted by some as a proof of a need for describing quantum mechanical processes in terms of the two-state vector formalism (TSVF). We are unable to understand the nature of interference in a Mach-Zehnder Interferometer (MFI) due to complementarity principle \cite{COMP}: the better we know the path of each photon passing MFI, the less likely it is to generate fringes at the output. The Authors claim that they have been able to put path markers on the light and yet maintain the interference. The markers were put in the form of oscillating mirrors of MFI (each with a different frequency), and rather than single photon detectors, a quad-cell detector (QCD) is used at one output. It yields time-resolved difference between light intensities on upper and lower parts of the detector. The output signal is then Fourier analyzed, with clearly distinguishable peaks at frequencies, at which MFI mirrors oscillate. These are, for the Authors, certificates of light interacting with particular mirrors. 

The central point of Ref. \cite{DANAN} is Figure 2(b). Therein, we see  a (big) MFI with another (small) MFI nested in the upper arm of the big one. Mirror C bends the lower arm of the big interferometer, while mirrors E an F direct light though an upper arm so that it passes the small one, based on mirrors A and B, and recombines with a beam from the lower arm. The big MFI is said to be tuned for a constructive interference at the output with QCD, however, the small loop is set for destructive interference at the output, which recombines with the lower arm. In such a case, the power spectrum has three distinct peaks, corresponding to frequencies $f_A, f_B,$ and $f_C$.

The Authors explain that the easiest way to explain these peaks is through TSVF. Namely, we should consider an overlap between two wavefunctions: evolved in time from the source and retracted from the detector. Due to fine tuning of the small MFI, they overlap on three parts: the lower path containing mirror C, and both arms of the small MFI, with mirrors A and B. In talks on this work, Vaidman claims that this result certifies each photon to have been in all three places.

The alternative explanation of this effect, based on wave optics, is widely discussed \cite{comment}. The key point of this critism is that it is impossible to move mirrors with maintaining the desired alignment of the interferometers. If we assume that a piezoelectric element moving the mirror is placed 1 cm from an axis and oscillates with amplitude of 10 nm, the light spot is displaced by 100 nm, give that the length between the mirror and a beam splitter is 10 cm. This is enough to change the interference patterns, even without instruments as sensitive as those used to observe occultations of dim stars by exoplanets. Indeed, the Authors admit that they were able to achieve about $95\%$ visibility.

Here, I want to raise another issue concerning the TSVF interpretation. I particularly focus on Figure 2(c), where the lower arm is blocked. At first, it seems that the power spectrum is flat with zero power reaching QCD. This is acceptable, since there is a block behind mirror C and the small MFI is tuned for destructive interference. Yet, a close inspection reveals some structure of the spectrum. We see some peaks at about 280 and 310 Hz, which are visible on other figures and hence are likely to correspond to some acoustic eigenmodes of the setup. If the Authors were asked to present their spectra normalized to a fixed integral over range 270-340 Hz, the most distinct peak would be at $f_E$. If the TVSF interpretation was correct, all the light in the function retracted from the detector would leave the setup through a beam splitter placed between mirrors A, B, and E, so the two vectors would not overlap on $E$. In such a case, we should rather observe peaks on $f_A$ and $f_B$, which are highly questionable in Figure 2(c). On the other hand, the classical and quantum field theories easily explain why $f_E$ is the dominant peak: it simply comes from the fact that E is the farthest mirror from the detector. 

In conclusion, Ref. \cite{DANAN} does not describe any effect that cannot be explained by any accepted physical theories. In fact, the new formalism is in disagreement with the data presented in the Letter. Certainly, the data supplied by the Authors are insufficient to draw conclusions about correctness of the present theory or multilocation of particles. Currently there is no effect known, which could be explained only by TVSF, and  hence it should be cut off by the Ockham's razor.

{\bf Marcin Wie\'sniak} Institute of Theoretical Physics and Astrophysics, University of Gda\'nsk, Poland. Work financed by the European Comission as a part of project BRISQ2. The work is subsidized form funds for science for years 2012-2015 
approved for
international co-financed project BRISQ2 by Polish Ministry of Science 
and Higher Education (MNiSW).

\end{document}